\documentclass[12pt]{article}

\usepackage{epsfig}

\def\square{\kern1pt\vbox{\hrule height 1.2pt
\hbox{\vrule width 1.2pt\hskip 3pt
\vbox{\vskip 6pt}\hskip 3pt\vrule width 0.6pt}
\hrule height 0.6pt}\kern1pt}
\def\ltwid{\mathrel{\raise.3ex\hbox{$<$\kern-.75em\lower1ex\hbox{$\sim$}}}}
\def\gtwid{\mathrel{\raise.3ex\hbox{$>$\kern-.75em\lower1ex\hbox{$\sim$}}}}

\begin{document}

\begin{titlepage}
\begin{flushright}
CCTP-2013-09 \\ UFIFT-QG-13-06
\end{flushright}

\vspace{0.5cm}

\begin{center}
\bf{Pure Gravitational Back-Reaction Observables}
\end{center}

\vspace{0.3cm}

\begin{center}
N. C. Tsamis$^{\dagger}$
\end{center}
\begin{center}
\it{Institute of Theoretical \& Computational Physics, and \\
Department of Physics, University of Crete \\
GR-710 03 Heraklion, HELLAS.}
\end{center}

\vspace{0.2cm}

\begin{center}
R. P. Woodard$^{\ast}$
\end{center}
\begin{center}
\it{Department of Physics, University of Florida \\
Gainesville, FL 32611, UNITED STATES.}
\end{center}

\vspace{0.3cm}

\begin{center}
ABSTRACT
\end{center}
\hspace{0.3cm} 
After discussing the various issues regarding and requirements 
on pure quantum gravitational observables in homogeneous-isotropic
conditions, we construct a composite operator observable satisfying 
most of them. We also expand it to first order in the loop counting 
parameter and suggest it as a physical quantifier of gravitational 
back-reaction in an initially inflating cosmology. 

\vspace{0.3cm}

\begin{flushleft}
\end{flushleft}

\vspace{0.1cm}

\begin{flushleft}
$^{\dagger}$ {\it e-mail:} tsamis@physics.uoc.gr \\
$^{\ast}$ {\it e-mail:} woodard@phys.ufl.edu
\end{flushleft}

\end{titlepage}

\section{Introduction}

Because the quantum gravitational back-reaction on inflation is 
contentious it is worth reviewing the mechanism of back-reaction 
in general terms:
\\ [3pt]
$\bullet$ The Uncertainty Principle requires every degree of
freedom of every dynamical variable to experience a minimal 
level of excitation known as its 0-point motion;
\\ [3pt]
$\bullet$ 0-point motions are affected by background force 
fields; and
\\ [3pt]
$\bullet$ 0-point motions contribute to the currents and 
stress-energies which source background force fields.
\\ [3pt]
The classic example is establishing a uniform electric field, 
which leads to the production of charged particle pairs 
\cite{Schwinger}. The resulting currents inevitably change 
the electric field that caused them \cite{Mottola,SPad,TTW}.

That said, there are two controversial issues regarding the extent
to which quantum fluctuations can really change macroscopic force
fields:
\\ [3pt]
$\bullet$ Distinguishing physical sources from gauge artifacts; 
and
\\ [3pt]
$\bullet$ Distinguishing changes in background force fields from 
changes in the distribution of quantum fluctuations.
\\ [3pt]
While the reality of back-reaction in quantum electrodynamics is
established, this is not the case about analogous back-reaction 
potential effects to particle production in an expanding universe 
\cite{Sasha,Emil,TW1,Robert}. Undoubtedly, quantum gravity provides 
fertile soil for such concerns because it has so far defied an 
accepted consistent perturbative formulation \cite{RPW}, because 
it possesses no local gauge invariant observables \cite{TW2}, and
because even the simplest tree order amplitudes are difficult to
compute \cite{Sigurd}. On the other hand, it is counter-productive 
to generically criticize any result, for instance, as either a 
gauge artifact or an inappropriate choice of vacuum \cite{GT}.

That is not to deny the validity of concerns about the reality 
of quantum gravitational back-reaction. As an example, consider
the assertion that infrared effects in scalar-driven inflation 
induce a secular slowing of the expansion rate at 1-loop order 
\cite{ABM}. Unruh quickly raised the issue of the gauge independence 
of the effect \cite{Unruh}. The secular slowing was found in other 
simple gauges as well \cite{AW1}. However, all secular 1-loop 
effects disappeared when a fully invariant technique was employed 
for fixing the surface of simultaneity based on the state of the 
inflaton field \cite{AW2,BG1}. Subsequent work on the same problem 
has demonstrated that different results can emerge when other fields 
are employed as the clock variable \cite{BG2,FMVV,MV1,MVB,MV2}.

The standard local definition of the expansion rate 
${\cal H}$ \cite{HawkEll}:
\footnote{Hellenic indices take on spacetime values 
while Latin indices take on space values. Our metric
tensor $g_{\mu\nu}$ has spacelike signature.}
\begin{equation}
{\cal H}(t, {\bf x})
\, = \, 
\frac13 \, D^{\mu} u_{\mu}(t, {\bf x})
\;\; , \label{H}
\end{equation}
is in terms of the covariant derivative $D_{\mu}$ of 
a timelike 4-velocity field $u_{\mu}$:
\begin{equation}
g^{\mu\nu}(x) \, u_{\mu}(x) \, u_{\nu}(x) 
\, = \, - 1
\;\; . \label{u}
\end{equation}
From the integral curves of the 4-velocity field we can
determine whether the universe expands or contracts by
showing whether these integral curves further diverge 
or converge, respectively. However, much depends upon 
the choice of the 4-velocity field. For example, even 
in classical de Sitter, with no quantum effects at all, 
it is possible to get the local expansion rate to be 
positive or negative by choosing $u_{\mu}$ to be the 
gradient of the time variable on either open or (early) 
closed coordinates, respectively \cite{TW3}!

A standard choice for the 4-velocity in scalar-driven 
inflation is to use the scalar inflaton field $\phi$ to 
construct the 4-velocity $u_{\mu}$ \cite{BG1}:
\begin{equation}
u_{\mu} \, \equiv \,
- \frac{\partial_{\mu} \phi}
{\sqrt{-g^{\alpha\beta} \; \partial_{\alpha} \phi \;
\partial_{\beta} \phi}}
\;\; . \label{u_infl}
\end{equation}
This 4-velocity field is not in general timelike. However,
there is no problem in perturbation theory and while $\phi$ 
is rolling down its potential, provided that the change 
in its classical value per time interval is larger than 
its quantum fluctuation in the same time interval. Moreover, 
by expanding about the classical inflaton ${\bar \phi}(t)$
in a background $FRW$ geometry:
\begin{equation}
\phi(t, {\bf x}) \, = \,
{\bar \phi} (t) \, + \, \delta \phi(t, {\bf x})
\;\; , \label{phi}
\end{equation}
we can fix the time $t$ by requiring \cite{BG1}:
\begin{equation}
\delta \phi(t, {\bf x}) \, = \, 0
\;\; . \label{phi_time}
\end{equation}
Then, the 4-velocity field (\ref{u_infl}) just described 
corresponds to the field of observers co-moving with the 
inflaton. 

Our concern in this paper is not scalar-driven inflation 
but rather the back-reaction that would occur in pure 
quantum gravity if the universe was released in a prepared 
state that is initially locally de Sitter with Hubble 
parameter $H_I$ \cite{TW4}. In this case there is no scalar 
inflaton to furnish a clock but we shall see that it is 
possible to construct a number of non-local scalar functionals 
of the metric which measure the elapsed time from the initial 
value surface. Any of these can be used to define a timelike 
4-velocity field $u_{\mu}$ and to fix the surface of simultaneity, 
exactly as is done in scalar-driven inflation. Fixing the space 
point invariantly is no more possible for us than it is in 
scalar-driven inflation \cite{TU}, but this is not considered 
problematic for scalar-driven inflation and it should be alright 
for pure gravity provided the initial state is homogeneous and 
isotropic.
\footnote{Ultimately, it is the cosmological principle which
precludes the existence of any special space point on which 
to define an observable. It is only when non-homogeneous and 
isotropic sources are present that we can use their location 
as space points on which to observe. However, the presence of 
an initial spacelike surface allows us to fix time physically
and not by enforcing a gauge condition. In one sentence: 
{\it time can be invariantly determined but space cannot}.}

\section{A Gravitational Geometrical Observable} 

In the case of pure gravity there is no matter field present
-- like the inflaton -- to be used as a time clock. The physical
conditions we shall investigate assume -- at some initial time
$t_I$ -- a prepared initial homogeneous and isotropic state 
and an inflating universe approximated by the de Sitter geometry.
\footnote{Hereafter, we shall work in $D$ spacetime dimensions
to eventually facilitate dimensional regularization. In open 
coordinates, otherwise known as the cosmological patch, our de 
Sitter line element is: $ds^2 = -dt^2 + a^2(t) \, d{\bf x} \cdot 
d{\bf x} = -dt^2 + \exp(2 H_I t) \, d{\bf x} \cdot d{\bf x}$ ,
where $H_I$ is the (constant) Hubble parameter.}
The invariant volume of the past light cone ${\cal V}$ is a 
geometrical object that can be used by an observer as his time 
clock. At the observation point $x$ it equals \cite{PW}:
\begin{equation}
{\cal V}[g] (x) \, \equiv \;
\int d^D x' \; \sqrt{-g(x')} \; \theta(-\ell^2(x ; x'))
\;\; , \label{Vplc}
\end{equation}
where to find the length $\ell^2(x ; x')$ we construct 
the geodesic $\chi^{\mu}$ from $x'$ to $x$:
\begin{eqnarray}
& \mbox{} &
{\ddot \chi}^{\mu}(\tau) \, + \,
\Gamma^{\mu}_{~\rho\sigma}[\chi(\tau)] \;
{\dot \chi}^{\rho}(\tau) \;  
{\dot \chi}^{\sigma}(\tau) \, = \, 0 
\;\; , \\ 
& \mbox{} &
\chi^{\mu}(0) = x'^{\mu}
\,\, , \,\,
\chi^{\mu}(1) = x^{\mu}
\;\; , \label{geodesic}
\end{eqnarray}
and then use it to obtain the desired length:
\begin{equation}
\ell^2(x ; x') \, = \,
g_{\mu\nu}(x) \; {\dot \chi}^{\mu}(1) \; {\dot \chi}^{\nu}(1)
\;\; . \label{length}
\end{equation}
Since the volume of the past light cone is a monotonically 
increasing function of time, it can serve as a geometrically
meaningful time clock from which to construct the timelike 
$D$-velocity field thusly:
\begin{equation}
v_{\mu}[g](x) \, \equiv \,
- \frac{\partial_{\mu} {\cal V}[g](x)}
{\sqrt{-g^{\alpha\beta} \; 
\partial_{\alpha} {\cal V}[g](x) \; 
\partial_{\beta} {\cal V}[g](x)}}
\;\; . \label{v}
\end{equation}
We can invariantly fix the observation time by specifying 
the surfaces of simultaneity:
\begin{equation}
{\cal V}[g] \Big( T[g](x), {\bf x} \Big) 
\, = \,
\bar{\cal V}(t) 
\;\; , \label{T}
\end{equation}
where $\bar{\cal V}(t)$ is the volume of the past light cone 
in de Sitter spacetime. This requirement determines the 
functional $T[g](x)$ or, equivalently, the observation
time.

The expansion variable is given by (\ref{H}):
\begin{equation}
{\cal H}[g](x)
\, = \, 
\frac{1}{D-1} \, D^{\mu} v_{\mu}[g](x)
\, = \,
\frac{1}{D-1} \, \frac{1}{\sqrt{-g \,}} \,
\partial_{\mu} \Big( \sqrt{-g} \, g^{\mu\nu} v_{\nu} \Big)
\;\; , \label{H2}
\end{equation}
and its time can be invariantly fixed by inheriting (\ref{time})
to form:
\begin{equation}
{\rm H}[g](x) \, = \,
{\cal H}[g] \Big( T[g](x), {\bf x} \Big)
\; \; . \label{H2time}
\end{equation}
The expansion variable ${\rm H}$ does not have its space
position invariantly fixed; indeed this is impossible in
pure gravity on homogeneous and isotropic backgrounds. 
A completely invariant observable can be generated by
integrating ${\cal H}$ over the spacetime manifold taking
into account condition (\ref{T}):
\begin{equation}
\int d^D x \; \sqrt{-g(x)} \times {\cal H}(x) \times
\delta \Big[ {\cal V}[g]\Big( T[g](x), {\bf x} \Big) 
- \bar{\cal V}(t) \Big]
\;\; . \label{H2inv}
\end{equation}
While the latter is fully invariant, it may not be the 
most appropriate object to consider for our physical
problem; the necessary presence of $\sqrt{-g}$ in
(\ref{H2inv}) with its lack of derivatives would make
it the dominant contribution, suppressing the physical
results imprinted in ${\cal H}$. It is therefore 
preferable to use ${\rm H}$ -- given by (\ref{H2time})
-- as our observable; in spite of not being fully 
invariant, it is a scalar at an invariantly fixed time. 
And as mentioned earlier, it is not possible to invariantly 
fix space in a homogeneous and isotropic universe.

We wish to perturbatively calculate the expectation value 
of ${\rm H}$ in the presence of a homogeneous and isotropic 
initial state, in an initially inflating universe with 
Hubble parameter $H_I$. In such a setup it is desirable 
to conformally re-scale the metric:
\footnote{The relation between co-moving $t$ and conformal
$\eta$ times is: $dt = a(t) \, d\eta$. The de Sitter scale 
factor in conformal coordinates is $a(\eta) = - (H_I \, 
\eta)^{-1}$. The fluctuating graviton field is $h_{\mu\nu}
(\eta, {\bf x})$ and its trace $h(\eta, {\bf x})$. 
Differentiation(s) with respect to $\eta$ shall be indicated 
with prime(s).}
\begin{equation}
g_{\mu\nu} \, = \,
a^2 \, {\widetilde g}_{\mu\nu} \, = \, 
a^2 \Big( \eta_{\mu\nu} + \kappa h_{\mu\nu} \Big)
\;\; , \label{g_conf}
\end{equation}
since such a re-scaling preserves the sign of the length
$\ell^2(x ; x')$; null geodesics remain the same while 
spacelike and timelike geodesics remain spacelike and 
timelike respectively.

It is a cumbersome but straightforward exercise to expand
the basic elements comprising ${\rm H}$ in powers of the
parameter $\kappa$ starting from the basic expansion 
(\ref{g_conf}) of the metric in terms of the de Sitter 
background and the fluctuating graviton field $h_{\mu\nu}$:
\begin{eqnarray}
T[g](x) &\!\! = \!\!&
\eta \, + \, \kappa T_1 (\eta, {\bf x}) \, + \, 
\kappa^2 \, T_2 (\eta, {\bf x}) \, + \, \dots
\;\; , \label{pertT} \\
{\cal V}[g](x) &\!\! = \!\!&
\bar{\cal V}(\eta) \, + \, 
\kappa {\cal V}_1 (\eta, {\bf x}) \, + \, 
\kappa^2 \, {\cal V}_2 (\eta, {\bf x}) \, + \, \dots
\;\; , \label{pertV} \\
v_{\mu}[g](x) &\!\! = \!\!&
- a \Big( \delta_{\mu}^{~0} \, + \, 
\kappa v_{\mu \, 1} (\eta, {\bf x}) \, + \, 
\kappa^2 \, v_{\mu \, 2} (\eta, {\bf x}) \, + \, \dots \Big)
\;\; , \label{pertv} \\
{\cal H}[g](x) &\!\! = \!\!&
H_I \, + \, \kappa {\cal H}_1 (\eta, {\bf x}) \, + \,
\kappa^2 \, {\cal H}_2 (\eta, {\bf x}) \, + \, \dots
\;\; . \label{pertH}
\end{eqnarray}

However, a detailed examination of the 1-loop contributions
to the expectation value revealed the presence of undesirable
divergences. These divergences are not ultraviolet because 
they do not occur at coincident points. They occur because 
the propagator between two points on the {\it same} light 
ray diverges. It is unknown how to handle such infinities and 
therefore we shall dispense with the use of this particular
geometrically motivated observable as an indicator of 
back-reaction.

\section{A Gravitational Dynamical Observable}

Perhaps a more generic -- but also calculationally accessible 
-- observable can be constructed by considering a scalar 
functional $\Phi$ of the metric satisfying, for all $x$, the 
dynamical equation:
\begin{eqnarray}
\square \Phi[g](x) &\!\! = \!\!& (D-1) H_I
\;\; , \label{Phi} \\
&\!\! = \!\!&
\frac{1}{\sqrt{-g}} \,
\partial_{\mu} [\sqrt{-g} \, g^{\mu\nu} \, \partial_{\nu} \Phi]
\, = \,
\frac{1}{a^D \sqrt{-{\widetilde g}}} \,
\partial_{\mu} \Big[ a^{D-2} \sqrt{-{\widetilde g}} \, 
{\widetilde g}^{\mu\nu} \, \partial_{\nu} \Phi \Big]
\; . \qquad \label{Phi2}
\end{eqnarray}
In order to solve (\ref{Phi}) for the scalar $\Phi$, 
we must supply two conditions on the initial value 
surface (IVS):
\begin{equation}
\Phi(\eta_I, {\bf x}) \Big\vert_{\rm IVS} \, = \, 0
\quad , \quad
-g^{\alpha\beta}(\eta_I, {\bf x}) \; 
\partial_{\alpha} \Phi(\eta_I, {\bf x}) \;
\partial_{\beta} \Phi(\eta_I, {\bf x}) 
\Big\vert_{\rm IVS} \, = \, 1
\;\; . \label{IVD}
\end{equation}
By the aforementioned procedure, the $D$-velocity field 
$V_{\mu}$ is:
\footnote{In general, the $D$-velocity field $V_{\mu}$
is not timelike. It is in perturbation theory and for
the class of cosmollgical spacetimes of interest.}
\begin{equation}
V_{\mu} [g](x) \, \equiv \,
- \frac{\partial_{\mu} \Phi[g](x)}
{\sqrt{-g^{\alpha\beta}(x) \; \partial_{\alpha} \Phi[g](x) \;
\partial_{\beta} \Phi[g](x)}}
\;\; , \label{4-vel}
\end{equation}
and the expansion variable according to (\ref{H}) is: 
\begin{equation}
{\cal H}[g](x) \, = \,
\frac{1}{D-1} \, D^{\mu} V_{\mu}[g](x)
\, = \,
\frac{1}{D-1} \, \frac{1}{\sqrt{-g}} \,
\partial_{\mu} [\sqrt{-g} \, g^{\mu\nu} \, V_{\nu}]
\;\; . \label{H3}
\end{equation}
The surfaces of simultaneity are defined in a way analogous
to (\ref{T}):
\begin{equation}
\Phi[g](\vartheta [g](x), {\bf x}) \, = \,
\bar\Phi (\eta)
\;\; , \label{time}
\end{equation}
and they fix time in an invariant way. Our observable 
${\rm H}$ -- which physically represents the expansion 
rate of spacetime -- is given by :
\begin{equation}
{\rm H}[g](x) \, \equiv \,
{\cal H}[g](\vartheta [g](x), {\bf x})
\;\; . \label{H3time}
\end{equation}
Under general coordinate transformations, the variables
just constructed transform as follows:
\begin{equation}
{\cal H}[g'] (x) \, = \, {\cal H}[g] (x'^{\, -1}(x))
\quad , \quad
{\rm H}[g'] (\eta, {\bf x}) \, = \, 
{\rm H}[g] (\eta, x'^{\, -1}(\eta, {\bf x}))
\;\; . \label{transf}
\end{equation}

In order to perturbatively compute the expectation value of 
(\ref{H3time}) we should expand it in the parameter $\kappa$. 
The relevant expansions are given by (\ref{g_conf}, 
\ref{pertT}-\ref{pertH}) with the understanding that we must 
effect the trivial replacements of $(T, {\cal V}, v)$ with 
$(\vartheta, \Phi, V)$ respectively. 

As a first step, it is the 1-loop result we shall be interested. 
Thus, we need to find the expansion of ${\rm H}$ to order 
$\kappa^2$. Starting from the expansion of the scalar $\Phi$: 
\begin{equation}
\Phi \, = \,
\bar\Phi + \kappa \Phi_1 + \kappa^2 \Phi_2 + \dots
\;\; . \label{pertPhi}
\end{equation}
and substituting it in (\ref{Phi}-\ref{IVD}) we arrive 
at the following equations to lowest order in $\kappa$:
\footnote{The equation of motion (\ref{Phi}) implies: \\
$(D-1) H_I = \square{\Phi} = \bar{\square}\Phi - 
a^{-D} \, \partial^{\mu} [ a^{D-2} \, \kappa h_{\mu\rho} \,
{\widetilde g}^{\rho\nu} \, \partial_{\nu} \Phi ] +
2\kappa a^{-2} h_{\rho\sigma , \mu} \, 
{\widetilde g}^{\rho\sigma} \, {\widetilde g}^{\mu\nu} \,
\partial_{\nu} \Phi$ , \\
where we have defined: 
$\bar{\square} \equiv 
a^{-D} \, \partial_{\mu} 
[ a^{D-2} \eta^{\mu\nu} \partial_{\nu} ] =
a^{-2} \, [ - \partial_0^2 - (D-2) Ha \, \partial_0 + \nabla^2 ]$.}
\begin{equation}
\frac{1}{a^D} \; \partial_{\mu}
\Big[ a^{D-2} \, \partial_{\mu} \bar\Phi \Big] 
\, = \, (D-1) H_I
\quad , \quad
\bar\Phi \Big\vert_{\rm IVS} \, = \, 0 
\;\; , \label{Phi_0eqn}
\end{equation}
and to first order in $\kappa$:
\begin{eqnarray}
\frac{1}{a^D} \; \partial_{\mu}
\Big[ a^{D-2} \, \partial^{\mu} \Phi_1 \Big] 
&\!\! = \!\!& 
\frac{1}{a^D} \left\{ 
\partial_{\mu} \Big[ a^{D-2} \, h^{\mu\nu} \partial_{\nu} \bar\Phi \Big]
- \frac12 \, a^{D-2} \, h_{, \mu} \, \partial^{\mu} \bar\Phi \right\}
\;\; , \qquad \label{Phi_1eqn} \\
\Phi_1 \Big\vert_{\rm IVS} &\!\! = \!\!& 0
\quad , \quad
\frac{1}{a} \, \Phi'_1 \Big\vert_{\rm IVS} \, = \, 
\frac12 \, h_{00} \Big\vert_{\rm IVS}
\;\; . \label{Phi_1ivd}
\end{eqnarray}
The solution to (\ref{Phi_0eqn}) for the lowest order term is:
\begin{equation}
\bar\Phi (\eta) \, = \, - \frac{\ln a}{H_I}
\quad \Longrightarrow \quad
\partial_{\mu} \bar\Phi \, = \, - \delta_{\mu}^{0} \, a
\;\; . \label{Phi_0}
\end{equation}
The corresponding solution to (\ref{Phi_1eqn}-\ref{Phi_1ivd}) 
is more complicated:
\begin{eqnarray}
\Phi_1(x) &\!\! = \!\!&
\int d^D x' \; G_A (x; x') \left\{ 
- \partial'^{\mu} \Big[ a'^{D-1} \, h_{\mu 0}(x') \Big] 
+ \frac12 \, a'^{D-1} \, h'(x') \right\}
\nonumber \\
& \mbox{} &
- \int_{\eta' = \eta_0} d^{D-1} x' \; G_A (x; x') \,
\frac12 \, h_{00}(x')
\;\; , \label{Phi_1}
\end{eqnarray}
where the Green's function $G_A$ satisfies:
\footnote{The analytic form of $G_A$ can be found in \cite{Onemli},
equation (5).}
\begin{equation}
{a^D} \, \bar{\square} G_A (x; x') \, = \,
\delta^D (x-x') 
\;\; . \label{box0}
\end{equation}
There is no need to explicitly solve for the second order term 
$\Phi_2$. Although it will appear in the ${\cal O}(\kappa^2)$
terms of the expansion of (\ref{H3time}), it will not contribute 
when we take the expectation value of ${\rm H}$.

We must also ensure -- to the requisite order in $\kappa$ --
that condition (\ref{time}) defining the surfaces of simultaneity 
is satisfied: 
\begin{eqnarray}
\bar\Phi (\eta)
&\!\! = \!\!&
\Phi (\eta + \Delta\eta , {\bf x})
\label{pert_time1} \\
&\!\! = \!\!&
\bar\Phi (\eta + \Delta\eta) \, + \,
\kappa \, \Phi_1 (\eta + \Delta\eta , {\bf x}) \, + \,
\kappa^2 \, \Phi_2 (\eta + \Delta\eta , {\bf x}) \, + \,
\dots
\;\; . \qquad \label{pert_time2}
\end{eqnarray}
This is equivalent to perturbatively solving for the time 
component of a spacetime point which undoes the change in 
the surfaces of simultaneity under coordinate transformations.
The result is straightforward to obtain given that:
\begin {equation}
\bar\Phi (\eta + \Delta\eta) \, = \,
\bar\Phi (\eta) \, + \,
\bar\Phi' (\eta) \, \Delta\eta \, + \,
\frac12 \, \bar\Phi'' (\eta) \, \Delta\eta^2 \, + \,
\dots
\;\; . \label{pert_time3}
\end{equation}
From (\ref{pert_time1}-\ref{pert_time3}) and (\ref{Phi_0})
we conclude:
\begin{equation}
\Delta\eta \, = \, 
- \frac{\kappa \, \Phi_1}{\bar\Phi'} \, + \, {\cal O}(\kappa^2)
\, = \,
- \frac{\kappa (D-1) H_I}{a} \, \Phi_1 \, + \, {\cal O}(\kappa^2)
\;\; . \label{eta}
\end{equation}

It is also straightforward -- albeit much more tedious -- 
to expand ${\rm H}$ (\ref{H3time}) to ${\cal O}(\kappa^2)$. 
Perhaps we can summarize this undertaking as a 5-step 
procedure:
\\ [5pt]
$\bullet \,$ We expand, using (\ref{pertPhi}), the ratio:
\begin{equation}
\frac{\cal H}{H_I} \, = \,
\left[ 1 + \frac{g^{\mu\nu} \, \partial_{\mu} \Phi \partial_{\nu}}
{(D-1) H_I} \right]
\frac{1}{\sqrt{- g^{\alpha\beta} \, \partial_{\alpha} \Phi \,
\partial_{\beta} \Phi}}
\;\; . \label{step1}
\end{equation}
$\bullet \,$  In the resulting expansion we substitute the 
solution (\ref{Phi_0}) for the lowest order scalar $\bar\Phi$.
\\ [5pt]
$\bullet \,$ We shift by the coordinate transformation (\ref{eta})
to obtain the observable ${\rm H}$:
\begin{equation}
\frac{\rm H}{H_I} \, = \,
\frac{\cal H}{H_I} \, + \,
\kappa \, \frac{\Phi_1}{a} \, \frac{{\cal H}'}{H_I} \, + \,
{\cal O}(\kappa^3)
\;\; . \label{step3}
\end{equation}
$\bullet \,$ Starting from the equation:
\begin{equation}
\left[ 1 + \frac{1}{(D-1) \, H_I \, a} \, \partial_0 \right]
\left( \frac{\Phi'_i}{a} \right)
\, = \,
\frac{1}{(D-1) H_I} \Big[ - \bar{\square} \Phi_i +
\frac{\nabla^2}{a^2} \, \Phi_i \Big]
\;\; , \label{step4}
\end{equation}
(where $i=1,2$) we substitute in the expression which emerged 
from the previous step the following relations:
\begin{eqnarray}
\left[ 1 + \frac{1}{(D-1) H_I \, a} \, \partial_0 \right]
\left( \frac{\Phi'_1}{a} \right)
&\!\! = \!\!&
\Big[ 1 + \frac{1}{(D-1) H_I \, a} \, \partial_0 \Big] h_{00}
\nonumber \\
& \mbox{} &
\hspace{-2.2cm}
+ \, \frac{1}{(D-1)H_I} \, \Big\{ \, 
\frac{1}{2a} \, h' - \frac{1}{a} \, h_{0i,i} + 
\frac{\nabla^2}{a^2} \, \Phi_1 \, \Big\}
\;\; , \label{step4a} \\
\left[ 1 + \frac{1}{(D-1) H_I \, a} \, \partial_0 \right]
\left( \frac{\Phi'_2}{a} \right)
&\!\! = \!\!&
\Big[ 1 + \frac{1}{(D-1) H_I \, a} \, \partial_0 \Big]
\Big\{ \, h_{0\mu} \, \partial^{\mu} \Phi_1 -
h_{0\mu} \, h_0^{~\mu} \, \Big\}
\nonumber \\
& \mbox{} &
\hspace{-2.2cm}
+ \, \frac{1}{(D-1) H_I} \, \Big\{ \,
\frac{1}{2a^2} \, h^{, \mu} \, \partial_{\mu} \Phi_1 - 
\frac14 ( h^{\mu\nu} \, h_{\mu\nu} )' - 
\frac{1}{2a} \, h^{, \mu} \, h_{0\mu}
\nonumber \\
& \mbox{} &
\hspace{-3.2cm}
+ \, \partial_i \Big[
- \frac{1}{a^2} \, h_i^{~\mu} \, \partial_{\mu} \Phi_1 +
\frac{1}{a} \, h_i^{~\mu} \, h_{0\mu} +
\frac{1}{a^2} \, \partial_i \Phi_2 \Big] \, \Big\}
\;\; . \label{step4b} 
\end{eqnarray}
$\bullet \,$ After the above operation we act all derivatives 
$\partial'_{\mu}$ and then use the equations of motion (\ref{Phi})  
to substitute all second time derivatives of $\Phi_i$ with:
\begin{equation}
\Phi''_i \, = \,
-a^2 \, \bar{\square} \, \Phi_i \, - \,
(D-2) \, H_I \, a \, \Phi''_i \, + \,
\nabla^2 \Phi_i
\qquad (i = 1,2)
\;\; . \label{step5}
\end{equation}

Execution of the above 5-step process and some algebraic manipulations
result in the final answer for the observable. We first present it
in a suggestive compact notation:
\begin{equation}
\frac{\rm H}{H_I} \, = \,
1 \, + \, \kappa \, \frac{\rm H}{H_I} \Bigg\vert_1 
\, + \, \kappa^2 \, \left\{ \frac{\rm H}{H_I} \Bigg\vert_{2 \; hh}
+ \frac{\rm H}{H_I} \Bigg\vert_{2 \; h\Phi}
+ \frac{\rm H}{H_I} \Bigg\vert_{2 \; \Phi\Phi}
+ \frac{\rm H}{H_I} \Bigg\vert_{2 \; \partial} \right\}
\, + \, {\cal O}(\kappa^3)
\;\; , \label{final}
\end{equation}
and then identify each of the terms. The ${\cal O}(\kappa)$
contribution is:
\begin{equation}
\frac{\rm H}{H_I} \Bigg\vert_1 
\, = \,
\frac12 \, h_{00} \, + \, \frac{h'_{ii}}{2 (D-1) H_I \, a} 
+ \partial_i \Big[ - \frac{h_{0i}}{(D-1) H_I \, a} 
+ \frac{\partial_i \Phi_1}{(D-1) H_I \, a^2} \Big]
\;\; , \label{final1} 
\end{equation}
Its expectation value requires the addition of a single 
interaction vertex to reach ${\cal O}(\kappa^2)$.
\footnote{These results are simple to extract from the graviton
1-point function \cite{TW5}.}
There are four kinds of ${\cal O}(\kappa^2)$ contributions: 
\begin{eqnarray}
\frac{\rm H}{H_I} \Bigg\vert_{2 \; hh} 
&\!\! = \!\!&
\frac{1}{(D-1) H_I \, a} \, \Bigg\{ \,
\frac38 (D-1) H_I \, a \, h_{00} \, h_{00}
- \frac12 (D-1) H_I \, a \, h_{0i} \, h_{0i}
\nonumber \\
& \mbox{} &
- \frac12 h_{ij} \, h'_{ij}
+ \frac12 h_{00 , i} \, h_{0i}
+ \frac14 h_{00} \, h'_{ii}
- \frac12 h_{jj , i} \, h_{0i} \Bigg\}
\;\; , \label{final2hh} \\
\frac{\rm H}{H_I} \Bigg\vert_{2 \; h\Phi} 
&\!\! = \!\!&
\frac{1}{(D-1) H_I \, a^2} \, \Bigg\{ \,
\frac12 h_{00} \, \nabla^2 \Phi_1 
+ \frac12 (D-1) H_I \, a \, h'_{00} \, \Phi_1
\nonumber \\
& \mbox{} &
+ \frac12 \Big[ h''_{ii} - H_I \, a \, h'_{ii} - 
\nabla^2 h_{ii} \Big] \Phi_1 
+ \Big[ H_I \, a \, h_{0i , i} - h'_{0i , i} \Big] \Phi_1
\Bigg\}
\;\; , \label{final2hPhi} \\
\frac{\rm H}{H_I} \Bigg\vert_{2 \; \Phi\Phi} 
&\!\! = \!\!&
\frac{1}{2a^2} \, \frac{D+1}{D-1} \, (\partial_i \Phi_1)
(\partial_i \Phi_1)
\;\; , \label{final2PhiPhi} \\
\frac{\rm H}{H_I} \Bigg\vert_{2 \; \partial} 
&\!\! = \!\!&
\frac{1}{(D-1) H_I \, a} \; \partial_i \Bigg\{ \,
h_{i\mu} \, h_0^{~\mu} 
+ \frac12 h_{00} \, h_{0i}
- \frac{1}{a} h_{i\mu} \, \partial^{\mu} \Phi_1
+ \frac{1}{a} \partial_i \Phi_2
\nonumber \\
& \mbox{} &
- h_{00} \, \partial_i \Phi_1
+ \frac12 h_{jj} \, \partial_i \Phi_1 
- h_{0i} \, \Phi'_1
+ \partial_i \Big[ \Phi_1 ( \Phi'_1 - H_I \, a \, \Phi_1 ) \Big]
\Bigg\}
\label{final2partial}
\end{eqnarray}
The expectation values of these four operators involve only
propagators -- and no interaction vertices -- to reach 
${\cal O}(\kappa^2)$. Upon taking the expectation value 
of $\rm H$ in the presence of a spatially invariant state, 
total spatial derivatines will not contribute; for instance, 
all of (\ref{final2partial}) and the last term of (\ref{final1}).

\section{Epilogue}
An observable $\rm H$ has been constructed and expanded
to ${\cal O}(\kappa^2)$ for the purpose of quantifying 
the quantum gravitational back-reaction to an initially 
inflating universe by detecting changes to the expansion 
rate. Our observable is a non-local composite operator 
with an invariantly determined time. To compute its 
expectation value beyond 1-loop ($\kappa^2$) order it 
would be necessary also to include perturbative corrections 
to the initial state. These corrections take the form of 
new interactions on the initial value surface \cite{KOW}. 
However, because the lowest order corrections for quantum 
gravity are ${\cal O}(\kappa h^3)$, they cannot affect 
the expectation value of {\rm H} at ${\cal O}(\kappa^2)$.

However unphysical one might consider 1PI functions in a 
fixed gauge, they have a powerful advantage over the more 
invariant composite operators which are sometimes proposed 
to quantify back-reaction. This advantage is that conventional 
renormalization suffices to make non-coincident 1PI functions 
finite whereas composite operator renormalization is required 
for the more complicated operators. When these composite 
operators are not even local it is not understood how to 
carry out this renormalization.

Returning to our observable ${\rm H}$, its non-locality comes
from the scalar $\Phi$ which is determined from (\ref{Phi})
by inverting the diffferential operator $\square$. There is,
however, a way to make ${\rm H}$ local without altering 
physics in any way. We have so far been working with the 
pure gravity Lagrangian ${\cal L}_g$ and have introduced 
$\Phi$ via (\ref{Phi}). Except for the initial conditions
(\ref{IVD}) which do not affect the ultraviolet structure, 
we can introduce $\Phi$ on the Lagrangian level so that we 
obtain (\ref{Phi}) as its equation of motion:
\begin{eqnarray}
{\cal L}_g &\!\! = \!\!&
\frac{1}{16 \pi G} \, ( R - 2 \Lambda ) \sqrt{-g}
\quad \longrightarrow 
\label{Lg} \\
{\cal L}_{g+\Phi} &\!\! = \!\!&
{\cal L}_g \, - \,
\frac12 \, \partial_{\mu} \Phi \; \partial_{\nu} \Phi
\, g^{\mu\nu} \sqrt{-g} \, - \, 
(D-1) H_I \, \Phi \sqrt{-g} 
\;\; . \label{Lg+Phi}
\end{eqnarray}
The variable $\Phi$ -- which was a non-local functional 
$\Phi[g](x)$ of the metric in the theory (\ref{Lg}), is 
now a local field $\Phi(x)$ in the theory (\ref{Lg+Phi}). 
Thus, the $D$-velocity field $V_{\mu}$ (\ref{4-vel}) and the 
quantities ${\cal H}$, ${\rm H}$ (\ref{H3}, \ref{H3time})
automatically become local. The renormalization properties 
of local composite operators are understood \cite{SW,IZ}.

It remains to actually compute the expectation value of
${\rm H}$; first to 1-loop order \cite{kitamoto} -- where 
the renormalized correction should be zero -- and then, 
hopefully, to 2-loop order where the first signals of a 
secular back-reaction are expected \cite{TW1, TW4}.

\vspace{1cm}
%\newpage

\centerline{\bf Acknowledgements}

We thank H. Kitamoto for extensive discussions on the 
problems of computing the expectation value of our 
observable at 1-loop order.
This work was partially supported by European Union program 
Thalis ESF/NSRF 2007-2013 MIS-375734, by European Union 
(European Social Fund, ESF) and Greek national funds through
the Operational Program “Education and Lifelong Learning” of 
the National Strategic Reference Framework (NSRF) under 
“Funding of proposals that have received a positive evaluation 
in the 3rd and 4th Call of ERC Grant Schemes”, by NSF grant 
PHY-1205591, and by the Institute for Fundamental Theory at 
the University of Florida.

\end{document}